# ABDUCTION WITH PENALIZATION IN LOGIC PROGRAMMING

**Giovambattista Ianni, Nicola Leone, Simona Perri**
Department of Mathematics
University of Calabria
I-87030 Rende (CS),
`ianni@deis.unical.it,`
`leone@unical.it,`
`sperri@si.deis.unical.it`

**Francesco Scarcello**
D.E.I.S.
University of Calabria
I-87030 Rende (CS), Italy
`scarcello@deis.unical.it`

**Abstract**

Abduction, first proposed in the setting of classical logics, has been studied with growing interest in the logic programming area during the last years.

In this paper we study *abduction with penalization* in logic programming. This form of abductive reasoning, which has not been previously analyzed in logic programming, turns out to represent several relevant problems, including optimization problems, very naturally. We define a formal model for abduction with penalization from logic programs, which extends the abductive framework proposed by Kakas and Mancarella. We show the high expressiveness of this formalism, by encoding a couple of relevant problems, including the well-know Traveling Salesman Problem from optimization theory, in this abductive framework. The resulting encodings are very simple and elegant. We analyze the complexity of the main decisional problems arising in this framework. An interesting result in this course is that "negation comes for free." Indeed, the addition of (even unstratified) negation does not cause any further increase to the complexity of the abductive reasoning tasks (which remains the same as for not-free programs).

## 1 Introduction

Abduction is an important kind of reasoning which has been first studied in depth by Peirce [18]. Given the observation of some facts, abduction aims at concluding the presence of other facts, from which, together with an underlying theory, the observed facts can be explained, i.e., deductively derived. Thus, roughly speaking, abduction amounts to an inverse of modus ponens.

For example, medical diagnosis is a typical abductive reasoning process: From the symptoms and the medical knowledge, a diagnosis about a possible disease is abduced. Notice that this form of reasoning is not sound (a diagnosis may turn out to be wrong), and that in general several abductive explanations (i.e., diagnoses) for observations may be possible.

During the last years, there has been increasing interest in abduction in different areas of computer science. It has been recognized that abduction is an important principle of common-sense reasoning, and that abduction has fruitful applications in a number of areas such diverse as model-based diagnosis [19], speech recognition [11], maintenance of database views [14], and vision [3].

In the past, most research on abduction concerned abduction from classical logic theories. However, we argue that the use of logic programming to perform abductive reasoning can be more appropriate in several applicative domains.

For instance, consider the following scenario. Assume that it is Saturday and is known that Joe goes fishing on Saturdays if it's not raining. This may be represented by the following theory $T$:

$$go\_fishing \leftarrow is\_saturday \ \& \ \neg rains \ ; \quad is\_saturday \leftarrow$$

Now you observe that Joe is not out for fishing. Intuitively, from this observation we conclude that it rains (i.e, we abduce $rains$), for otherwise Joe would be out for fishing. Nevertheless, under classical inference, the fact $rains$ is not an explanation of $\neg go\_fishing$, as $T \cup \{rains\} \not\models \neg go\_fishing$ (neither can one find any explanation). On the contrary, if we adopt the semantics of logic programming, then, according with the intuition, we obtain that $rains$ is an explanation of $\neg go\_fishing$, as it is entailed by $T \cup \{rains\}$.

In the context of logic programming, abduction has been first proposed by Kakas and Mancarella [13] and, during the recent years, common interest in the subject has been growing rapidly [4, 15, 13, 12, 6, 5, 20], also for the observation that, compared to deduction, this kind of reasoning has some advantages for dealing with incomplete information [5, 1].

In this paper we study *abduction with penalization* in logic programming. This form of abductive reasoning, well studied in the setting of classical logics [7], has not been previously analyzed in logic programming.

We define a formal model for abduction with penalization from logic programs, which extends the abductive framework proposed by Kakas and Mancarella [13]. Roughly, a penalty is assigned to each hypothesis. An abductive solution $S$ is weighted by the sum of the penalties of the hypotheses in $S$.[1] Mimimum-weight solutions are preferred over the other solutions since they are considered more likely to occur.

We show that some relevant problems, including, e.g., the classical *Traveling Salesman Problem* from optimization theory, can be encoded very simply and elegantly by abduction with penalization; while they cannot be encoded at all in (function-free) logic programming even under the powerful stable model semantics.

We analyze the computational complexity of the main decisional problems arising in this framework. An interesting result in this course is that "nega-tion comes for free." Indeed, the addition of (even unstratified) negation does not cause any further increase to the complexity of the abductive reasoning tasks (which remains the same as for not-free programs). Thus, abduction with penalization over general logic programs has precisely the same complexity as abduction with penalization over definite Horn theories of classical logics. Consequently, the user can enjoy the knowledge representation power of nonmonotonic negation without paying any additional cost in terms of computational overhead.

---
[1] Actually, in this paper, we consider also forms of weighting functions more general than Sum.

## 2 Preliminaries on Logic Programming

### 2.1 Syntax

A *term* is either a constant or a variable[2]. An *atom* is $a(t_1, ..., t_n)$, where $a$ is a *predicate* of arity $n$ and $t_1, ..., t_n$ are terms. A *literal* is either a *positive literal* $a$ or a *negative literal* $\neg A$, where $a$ is an atom. Two literals are *complementary* if they are of the form $p$ and $\neg p$, for some atom $p$. Given a literal $L$, $\neg.L$ denotes its complementary literal (the complement of an atom $A$ is literal $\neg A$ and vice versa). Accordingly, given a set $A$ of literals, $\neg.A$ denotes the set $\{\neg.L \mid L \in A\}$.

A *program clause* (or *rule*) $r$ is

$$a \leftarrow b_1, \cdots, b_k, \neg b_{k+1}, \cdots, \neg b_m, \qquad n \geq 1,\ m \geq 0$$

where $a, b_1, \cdots, b_m$ are atoms. Atom $a$ is the *head* of $r$, while the conjunction $b_1, ..., b_k, \neg b_{k+1}, ..., \neg b_m$ is the *body* of $r$.

A *(logic) program* is a finite set of rules. A $\neg$-free program is called *Horn program* or *positive program*. A term, an atom, a literal, a rule or program is *ground* if no variable appears in it. A ground program is also called a *propositional* program.

### 2.2 Stable model semantics

Let $LP$ be a program. The *Herbrand Universe* $U_{LP}$ of $LP$ is the set of all constants appearing in $LP$. The *Herbrand Base* $B_{LP}$ of $LP$ is the set of all possible ground atoms constructible from the predicates appearing in the rules of $LP$ and the constants occurring in $U_{LP}$ (clearly, both $U_{LP}$ and $B_{LP}$ are finite). Given a rule $r$ occurring in a program $LP$, a *ground instance* of $r$ is a rule obtained from $r$ by replacing every variable $X$ in $r$ by $\sigma(X)$, where $\sigma$ is a mapping from the variables occurring in $r$ to the constants in $U_{LP}$. We denote by $ground(LP)$ the (finite) set of all the ground instances of the rules occurring in $LP$. An *interpretation* for $LP$ is a subset $I$ of $B_{LP}$ (i.e., it is a set of ground atoms). A positive literal $a$ (resp. a negative literal $\neg a$) is true with respect to an interpretation $I$ if $a \in I$ (resp. $a \notin I$); otherwise it is false. A ground rule $r$ is *satisfied* (or *true*) w.r.t. $I$ if its head is true w.r.t. $I$ or its body is false

---
[2] Note that function symbols are not considered in this paper.

w.r.t. $I$. A *model* for $LP$ is an interpretation $M$ for $LP$ such that every rule $r \in ground(\mathcal{P})$ is true w.r.t. $M$.

Given a logic program $LP$ and an interpretation $I$, the *Gelfond-Lifschitz transformation* of $LP$ with respect to $I$ is the logic program $LP^I$ consisting of all rules $a \leftarrow b_1, \ldots, b_k$ such that (1) $a \leftarrow b_1, \ldots, b_k, \neg b_{k+1}, \ldots, \neg b_m \in LP$ and (2) $b_i \notin I$, for all $k < i \leq m$.

Notice that $\neg$ does not occur in $LP^I$, i.e., it is a positive program. Each positive program $LP$ has a least model (i.e., a model included in every model), denoted by $lm(LP)$.

An interpretation $I$ is called a *stable model* of $LP$ iff $I = lm(LP^I)$ [10]. The collection of all stable models of $LP$ is denoted by $\text{STM}(LP)$ (i.e., $\text{STM}(LP) = \{I \mid I = lm(LP^I)\}$).

**Example 2.1** Consider the following (ground) program $LP$:

$$a \leftarrow \neg b \qquad b \leftarrow \neg a \qquad c \leftarrow a \qquad c \leftarrow b$$

The stable models of $LP$ are $M_1 = \{a, c\}$ and $M_2 = \{b, c\}$. Indeed, by definition of Gelfond-Lifschitz transformation, $LP^{M_1} = \{\ a \leftarrow,\ c \leftarrow a,\ c \leftarrow b\ \}$ and $LP^{M_2} = \{\ b \leftarrow,\ c \leftarrow a,\ c \leftarrow b\ \}$; thus, it is immediately recognized that $lm(LP^{M_1}) = M_1$ and $lm(LP^{M_2}) = M_2$.

In general, a logic program may have more than one stable model or even no stable model at all. In the logic programming framework (under stable model semantics) there are two main notions of reasoning:

**Brave reasoning** (or *credulous reasoning*) infers that a literal $Q$ is true in $LP$ (denoted $LP \models^b Q$) iff $Q$ is true with respect to $M$ for some $M \in \text{STM}(LP)$.

**Cautious reasoning** (or *skeptical reasoning*) infers that a literal $Q$ is true in $LP$ (denoted $LP \models^c Q$) iff $Q$ is true with respect to $M$ for all $M \in \text{STM}(LP)$.

The inference relations $\models^b$ and $\models^c$ extend to sets of literals as usual.

**Example 2.2** For the program $LP$ of Example 2.1, $a, b$ and $c$ are brave inferences ($LP \models^b \{a, b, c\}$); the only cautious inference is $c$ ($LP \models^c c$).

In this paper, we are mainly interested in brave reasoning, even if our definitions can be easily extended to cautious reasoning.

## 3  A Model of Abduction with Penalization

In this section, we describe our formal model for abduction with penalizations over logic programs.

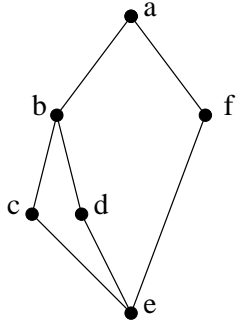
Figure 1: Computer network

**Definition 3.1** A problem of abduction $\mathcal{P}$ consists of a triple $\langle H, LP, O\rangle$, where $H$ is a finite set of ground atoms *(hypotheses)*, $LP$ is a logic program, and $O$ is a finite set of ground literals *(observations, or manifestations)*.

A set of hypotheses $S \subseteq H$ is an *admissible solution* (or *explanation*) to $\mathcal{P}$ if there exists a stable model $M$ of $LP \cup S$ such that, $\forall o \in O$, $o$ is true w.r.t. $M$ (i.e., $LP \cup S \models^b O$).

The set of all admissible solutions to $\mathcal{P}$ is denoted by $Adm(\mathcal{P})$. ∎

The following example shows a classical application of abduction for diagnosis purposes.

**Example 3.2** (NETWORK DIAGNOSIS) Suppose that we are working on machine $a$ (and we therefore know that machine $a$ is online) of the computer network in Figure 1, but we observe machine $e$ is not reachable from $a$, even if we are aware that $e$ is online. We would like to know which machines are offline. This can be easily modeled in our abduction framework defining a problem of abduction $\langle H, LP, O\rangle$, where the set of hypotheses is $H = \{$*offline(a), offline(b), offline(c), offline(d), offline(e), offline(f)*$\}$, the set of observations is $O = \{\neg$ *offline(a)*, $\neg$ *offline(e)*, $\neg$ *reaches(a,e)*$\}$, and $LP$ is the logic program

$reaches(X, X) \ :- \ node(X), \neg \ offline(X).$
$reaches(X, Z) \ :- \ reaches(X, Y), connected(Y, Z), \neg \ offline(Z).$

Note that the admissible solutions for $\mathcal{P}$ corresponds to the network configurations that may explain the observations in $O$. In this example, $Adm(\mathcal{P})$ contains five solutions $S_1 = \{$*offline(f), offline(b)*$\}$, $S_2 = \{$*offline(f), offline(c), offline(d)*$\}$, $S_3 = \{$*offline(f), offline(b), offline(c)*$\}$, $S_4 = \{$*offline(f), offline(b), offline(d)*$\}$, $S_5 = \{$*offline(f), offline(b), offline(c), offline(d)*$\}$.

Note that Definition 3.1 concerns only the logical properties of the hypotheses, and it does not take into account any kind of minimality criterion. We next define the problem of abduction with penalizations, which allows us to make finer abductive reasonings, by expressing preferences on different sets of hypotheses and single out the most plausible abductive explainations.

**Definition 3.3** A problem of abduction with penalization (*PAP*) $\mathcal{P}$ is a tuple $\langle H, LP, O, \gamma, cost\rangle$, where $\langle H, LP, O\rangle$, is a problem of abduction, $\gamma$ is a function from $H$ to the set of positive reals *(the penalty function)*, and $cost : 2^{R^+} \to R^+$ is a function computable in polynomial time. The set of admissible solutions for $\mathcal{P}$ is the same as the set of solutions of the embedded abduction problem $\langle H, LP, O \rangle$, i.e., we define $Adm(\mathcal{P}) = Adm(\langle H, LP, O \rangle)$.

For a set of atoms $A$, we denote by $\gamma(A)$ the set $\{\gamma(a) \mid a \in A\}$. Then, $S$ is an *(optimal) solution* (or *explanation*) for $\mathcal{P}$ if (i) $S \in Adm(\mathcal{P})$ and (ii) $cost(\gamma(S)) \leq cost(\gamma(S'))$, for all $S' \in Adm(\mathcal{P})$.

The set of all (optimal) solutions for $\mathcal{P}$ is denoted by $Opt(\mathcal{P})$. ∎

To simplify the notation, we will write $cost_\gamma$ to denote the composition of the cost function $cost$ with the penalty function $\gamma$, i.e., for a set of hypotheses $S$, $cost_\gamma(S) = cost(\gamma(S))$.

**Example 3.4** (MINIMUM-CARDINALITY CRITERION) Consider again the network and the problem of abduction $\langle H, LP, O \rangle$ in Example 3.2. From this problem, we define a problem of abduction with penalization $\mathcal{P} = \langle H, LP, O, \gamma, cost \rangle$, where the cost and the penalty functions are defined as follows. For each $S \subseteq H$, $cost_\gamma(S) = \sum_{x \in S} \gamma(x)$ and, for each $h \in H$, $\gamma(h) = 1$.

This way, we prefer the explanations with the minimum numbers of offline machines. Since $cost_\gamma(S_1) = 2$, $cost_\gamma(S_2) = cost_\gamma(S_3) = cost_\gamma(S_4) = 3$, $cost_\gamma(S_5) = 4$, it follows that $S_1$ is the unique optimal solution for $\mathcal{P}$.

**Example 3.5** (PROBABILISTIC CRITERION) Minimizing the number of offline machines is not necessarily the best strategy. If the reliability of the machines is very different, one should take it into account. To this end, for each machine $x$ in the network of Figure 1, the penalty function $\gamma$ assigns the probability of $x$ to be offline to the hypothesis *offline*$(x)$. Let these values be $\gamma(\textit{offline}(a)) = \gamma(\textit{offline}(e)) = 0.2$, $\gamma(\textit{offline}(b)) = 0.64$, $\gamma(\textit{offline}(f)) = 0.4$ and $\gamma(\textit{offline}(c)) = \gamma(\textit{offline}(d)) = 0.8$.

In this case, a good cost function should prefer solutions corresponding to the sets of machines that are most likely down simultaneously. We can reasonably assume these events to be independent each other. Thus, our cost function should prefer sets of machines such that the product of the probabilities (i.e., of the penalties) is the maximum. This is clearly equivalent to minimize the following cost function: $cost_\gamma(S) = 1 - \prod_{x \in S} \gamma(x)$, for any $S \subseteq H$. Thus, we have $cost_\gamma(S_1) = 0.74$, $cost_\gamma(S_2) = 0.74$, $cost_\gamma(S_3) = 0.80$, $cost_\gamma(S_4) = 0.80$, $cost_\gamma(S_5) = 0.84$. In this case, the optimal solutions for $\mathcal{P}$ are $S_1$ and $S_2$.

The following properties of a hypothesis in a *PAP* $\mathcal{P}$ are of natural interest with respect to computing abductive solutions.

**Definition 3.6** Let $\mathcal{P} = \langle H, LP, O, \gamma, cost \rangle$ be a *PAP* and $h \in H$. Then, $h$ is *relevant* for $\mathcal{P}$ iff $h \in S$ for some $S \in Opt(\mathcal{P})$, and $h$ is *necessary* for $\mathcal{P}$ iff $h \in S$ for every $S \in Opt(\mathcal{P})$.

**Example 3.7** In example 3.5, the only *necessary* hypothesis is *offline(f)*; while the *relevant* hypotheses are *offline(f)*, *offline(b)*, *offline(c)* and *offline(d)*.

The main decisional problems arising in the context of abduction with penalizations are the following. Given a *PAP* $\mathcal{P} = \langle H, LP, O, \gamma, cost \rangle$,

1. does there exist a solution for $\mathcal{P}$ ? (*Consistency*)

2. is a given set of hypotheses an optimal solution for $\mathcal{P}$? (*Optimality*)

3. is a given hypothesis $h \in H$ relevant for $\mathcal{P}$, i.e., does $h$ contribute to some optimal solution of $\mathcal{P}$? (*Relevance*)

4. is a given hypothesis $h \in H$ necessary for $\mathcal{P}$, i.e., is $h$ contained in all optimal solutions of $\mathcal{P}$? (*Necessity*)

The complexity of the consistency, optimality, relevance, and necessity problems is studied in Section 5.

## 4 An Example: The Traveling Salesman Problem

In [8], Eiter, Gottlob and Mannila show that Disjunctive Datalog (function-free logic programming with disjunction in the heads and negation in the bodies of the rules) is highly expressive, as it can express every problem in $\Sigma_2^P$ (that is, recognizable in polynomial time by a non-deterministic Turing machine which uses an NP oracle). Moreover, the authors strength the theoretical analysis of the expressiveness by proving that problems relevant in practice like, e.g., the *Traveling Salesman Problem (TSP)* and *Eigenvector*, can be programmed in Disjunctive Datalog, while they cannot be expressed by disjunction-free programs. Nevertheless, the logic programs implementing these problems in Disjunctive Datalog highlight, in our opinion, a weakness of the language for the representation of optimization problems. The programs are very complex and tricky, the language does not provide a clean and declarative way to implement these problems.[3]

We show below how the TSP problem can be encoded in our abductive framework. A comparison of our encoding against the encoding of this problem in (plain) disjunctive Datalog described in [8] clearly shows that abduction with penalizations provides a simpler, more compact, and more elegant encoding of TSP. Moreover, note that using this form of abduction even normal (disjunction-free) programs are sufficient for encoding such optimization problems.

---

[3]We refer to standard Disjunctive Datalog here. As shown in [2], the addition of *weak constraints*, implemented in the DLV system [9], is another way to enhance Disjunctive Datalog to naturally express optimization problems.

**Example 4.1** (TRAVELING SALESMAN BY ABDUCTION WITH PENALIZATIONS)
It is well-known that finding an optimal solution to the Traveling Salesman Problem (TSP) is intractable. Recall that the problem is, given cities $c_1, \ldots, c_n$, find a round trip that visits all cities in sequence and has minimal traveling cost, i.e., a permutation $\tau$ of $1, \ldots, n$ such that
$$w(\tau) = \sum_{i=1}^{n-1} w(\tau(i), \tau(i+1)) + w(\tau(n), \tau(1))$$
is minimum, where $w(i,j)$ is the cost of traveling from $c_i$ to $c_j$, given by an integer.

Computing an optimal tour is both NP-hard and co-NP-hard. In fact, in [16] it was shown that deciding whether the cost of an optimal tour is even is $\Delta_2^P$-complete. Hence, this is not possible in or-free logic programming even if unstratified negation is allowed (unless PH collapses).

Suppose that the cities are encoded by $1, \ldots, n$ and, for each pair of cities $1 \leq i, j \leq n$, The set of hypotheses is $H = \{C(i,j) \mid 1 \leq i, j \leq n\}$, where $C(i,j)$ encodes the fact that the salesman visits city $j$ immediately after city $i$. The penalty function $\gamma(C(i,j)) = w(i,j)$ encodes the cost of traveling from $j$ to $i$.

Let $LP$ be the program consisting of the following rules:[4]

(1)  $City(i) \leftarrow$                                    $1 \leq i \leq n$
(2)  $Visited(I) \leftarrow City(I), C(J,I), C(I,K)$
(3)  $MissedCity \leftarrow City(I), \neg Visited(I)$
(4)  $BadTour \leftarrow C(I,J), C(I,K), J \neq K$
(5)  $BadTour \leftarrow C(J,I), C(K,I), J \neq K$

Consider the PAP $\mathcal{P} = \langle H, \gamma, LP, O, cost \rangle$, where $O = \{\neg MissedCity, \neg BadTour\}$, and the cost function is the sum, i.e., $cost_\gamma(S) = \sum_{h \in S} \gamma(h)$, for any set of hypotheses $S \subseteq H$.

It is easy to see that every optimal solution $S \in Opt(\mathcal{P})$ corresponds to an optimal tour and vice versa. Intuitively, the facts (1) of $LP$ define the set of the cities to be visited. Rule (2) states that a city $i$ has been visited if it has been both reached and left by the traveling salesman (i.e., there exist two cities $j$ and $k$ such that $C(j,i)$ and $C(i,k)$ are in the solution). Rule (3) says that there is a missed city if at least one of the cities has not been visited. Atom $BadTour$, defined by rules (4) and (5), is true if some city is in two or more connection endpoints or connection startpoints. The observations $\neg MissedCity, \neg BadTour$ enforce that the tour is complete (no city is missed) and it is legal (no city is visited twice).

Thus, admissible solutions in $Adm(\mathcal{P})$ correspond one-to-one to the admissible (legal and complete) tours. Since optimal solutions minimize the sum of the connection costs, abductive solutions in $Opt(\mathcal{P})$ correspond one-to-one to the optimal tours. □

---

[4]The symbol $\neq$ represents the inequality predicate, we consider inequality a built-in without loss of generality, as it can be easily simulated in our formalism.

# 5 Computational Complexity

## 5.1 Preliminaries on Complexity Theory

For NP-completeness and complexity theory, cf. [17]. The classes $\Sigma_k^P, \Pi_k^P$ and $\Delta_k^P$ of the Polynomial Hierarchy (PH) are defined as follows:

$$\Delta_0^P = \Sigma_0^P = \Pi_0^P = \text{P} \quad \text{and for all } k \geq 1,$$
$$\Delta_k^P = \text{P}^{\Sigma_{k-1}^P}, \ \Sigma_k^P = \text{NP}^{\Sigma_{k-1}^P}, \ \Pi_k^P = \text{co-}\Sigma_k^P.$$

In particular, NP $= \Sigma_1^P$, co-NP $= \Pi_1^P$, and $\Delta_2^P = \text{P}^{\text{NP}}$. Here $\text{P}^C$ and $\text{NP}^C$ denote the classes of problems that are solvable in polynomial time on a deterministic (resp. nondeterministic) Turing machine with an oracle for any problem $\pi$ in the class $C$. The oracle replies to a query in unit time, and thus, roughly speaking, models a call to a subroutine for $\pi$ that is evaluated in unit time. If $C$ has complete problems, then instances of any problem $\pi'$ in $C$ can be solved in polynomial time using an oracle for any $C$-complete problem $\pi$, by transforming them into instances of $\pi$; we refer to this by stating that an oracle for $C$ is used. Notice that all classes $C$ considered here have complete problems.

The classes $\Delta_k^P$, $k \geq 2$, have been refined by the class $\Delta_k^P[O(\log n)]$, in which the number of calls to the oracle is in each computation bounded by $O(\log n)$, where $n$ is the size of the input. Notice that for all $k \geq 1$,

$$\Sigma_k^P \ \subseteq \ \Delta_{k+1}^P[O(\log n)] \ \subseteq \ \Delta_{k+1}^P \ \subseteq \ \Sigma_{k+1}^P \ \subseteq \ \text{PSPACE};$$

each inclusion is widely conjectured to be strict.

## 5.2 Complexity Results

We study the computational complexity of the main decision problems arising in the framework of abduction with penalization from logic programs. Throughout this section, we consider propositional *PAP*, i.e., we assume that the logic program of the *PAP* is ground.

**Theorem 5.1** (Consistency) *Deciding whether a PAP is consistent is* NP-*complete.*

**Proof.** Let $\mathcal{P} = \langle H, LP, O, \gamma, cost \rangle$ be a *PAP*.
(Membership). We can guess a set of hypotheses $S \subseteq H$ and a set of ground atoms $M$, and then check in polynomial time that (i) $M$ is a stable model of $LP \cup S$, and (ii) $O$ is true w.r.t. $M$.
(Hardness). Hardness can proved by noting that the problem of deciding if a logic program $LP$ has a stable model reduces to deciding consistency on the *PAP* problem $\langle \{\}, LP, \{\}, \gamma, cost \rangle$. ∎

**Theorem 5.2** (Optimality) *Deciding whether a set of atoms S is an optimal solution for a PAP $\mathcal{P}$ is* co-NP-*complete.*

**Proof.**(Membership). Let $\mathcal{P} = \langle H, LP, O, \gamma, cost \rangle$ be a *PAP* and let $S$ be a set of atoms. We can prove that $S$ is not an optimal solution for $\mathcal{P}$ by guessing a set of atoms $S' \subseteq H$ and then checking in polynomial time that (i) either $cost(\gamma(S')) < cost(\gamma(S))$ and $S' \in Adm(\mathcal{P})$, or (i) $S \notin Adm(\mathcal{P})$. It follows that deciding whether $S$ is optimal belongs to co-NP.

(Hardness). For the hardness part, recall the TSP encoding described in Example 4.1. This example shows how to construct, from an instance $T$ of the traveling salesman problem, a *PAP* $\mathcal{P}(T) = \langle H, LP, O, \gamma, cost \rangle$ such that the solutions of $T$ are in a one-to-one correspondence with the optimal solutions of $\mathcal{P}(T)$. Note that this construction is clearly feasible in polynomial time. Therefore, the co-NP-hard problem of deciding whether a tour for $T$ is a solution for $T$ (i.e., is a minimum-weighted tour) is (polynomial time) reducible to the problem of deciding whether a solution is optimal for a *PAP*. It follows that the optimality problem for a *PAP* is also co-NP-hard, and hence co-NP-complete.

**Theorem 5.3** *Let $\mathcal{P} = \langle H, LP, O, \gamma, cost \rangle$ be a PAP, where LP is a positive program. Deciding whether a set of atoms $S$ is an optimal solution for $\mathcal{P}$ is* co-NP-*complete.*

**Proof.** Consider the (non positive) program $LP$ of the *PAP* $\mathcal{P}(T)$ as in Example 4.1. Consider the positive program $LP'$ obtained from $LP$ by deleting rule (3), and let $O' = \{\neg BadTour\} \cup \{Visited(i) \mid 1 \leq i \leq n\}$ and $\mathcal{P}'(T) = \langle H, LP', O', \gamma, cost \rangle$. It is easy to verify that the solutions of $T$ are in a one-to-one correspondence with the optimal solutions of $\mathcal{P}'(T)$, and hence hardness holds even if negation cannot occur in the logic program of the *PAP*. ∎

Given a *PAP* $\mathcal{P} = \langle H, LP, O, \gamma, cost \rangle$, let $max\_cost(\mathcal{P})$ denote the maximum value that the function *cost* may return over all sets $H' \subseteq H$. In the following complexity results, we assume that $max\_cost(\mathcal{P})$ is polynomial-time computable from $\mathcal{P}$. Note that this is actually the case for all the considered examples, e.g., it is 1 for the network example. Moreover, it is always true if *cost* is a monotonic function, i.e., if $X_1 \subseteq X_2$ entails $cost_\gamma(X_1) \leq cost_\gamma(X_2)$. Indeed, in this case, $max\_cost(\mathcal{P}) = cost_\gamma(H)$. For instance, the sum function, classically used as the cost function in the framework of abduction with penalties, is clearly monotonic.

**Proposition 5.4** *Let $\mathcal{P} = \langle H, LP, O, \gamma, cost \rangle$ be a PAP, where $H$ and $O$ are propositional, LP is a propositional Horn program, $\gamma(h) = 1$ for each $h \in H$, and $cost(S) = \sum_{h \in S} \gamma(h)$. Define $PAP_{PR}$ (resp. $PAP_{PN}$) be the problem of deciding whether an hypothesis $h$ is relevant (resp. necessary) for $\mathcal{P}$. Then $PAP_{PR}$ (resp. $PAP_{PN}$) is $\Delta_2^P$-complete [7].*

**Theorem 5.5** (Relevance) *Deciding whether an hypothesis $h$ is relevant for a PAP $\mathcal{P}$ is $\Delta_2^P$-complete. Hardness holds even if the logic program of $\mathcal{P}$ is positive.*

**Proof.**(Membership). We next show that relevance is in $\Delta_2^P$. Let $\mathcal{P} = \langle H, LP, O, \gamma, cost \rangle$ be a *PAP* and let $h \in H$ be a hypothesis. We first compute the minimum value for the *cost* function over the set of admissible solutions for $\mathcal{P}$. We take advantage of an NP oracle $\mathcal{A}$ telling whether $\mathcal{P}$ has an admissible solution whose cost is less than a given value $s$, and of an NP oracle $\mathcal{B}$ telling whether there exists an admissible solution containing $h$ whose cost corresponds to a given value $s$. We proceed by binary search on $[0 \ldots k]$, where $k = max\_cost(\mathcal{P})$. We compute, by calling $\mathcal{A}$ at most $\log k$ times, the cost $c$ of the optimal solutions for $\mathcal{P}$. Then, we make a further call to $\mathcal{B}$ to determine whether there exists an admissible solution containing $h$ whose cost is $c$.[5]
(Hardness). It follows from Proposition 5.4. ∎

The complexity of the necessity problem can be proved in a very similar way.

**Theorem 5.6** (Necessity) *Deciding whether an hypothesis $h \in H$ is necessary for a PAP $\mathcal{P}$ is $\Delta_2^P$-complete. Hardness holds even if the logic program of $\mathcal{P}$ is positive.*

## 6 Further Examples

In this section we examine some further example where abduction turns out to be useful in order to represent real worlds problems.

**Example 6.1 (Strategic Companies).** In this example we propose a version of the classical *strategic companies* problem. The scenario of this problem includes a set of goods, and a set of companies which produce some good. Each company may own some quantity of shares of another company. The goal of the problem is to produce a given set of goods, by controlling directly or indirectly companies which produce the given set. To achieve the goal, companies can be bought, or ruled indirectly by controlling companies which own more than 50% of its shares. It is prescribed to minimize the quantity of money spent.

For simplicity of exposition, we will assume that, in order to control a company X, it always suffices to control at most two companies owning more than 50% of X's shares.

Companies configuration is encoded by introducing a ternary predicate symbol *share*, where $share(x, y, n)$ represents the fact that company $y$ is in possession of a quota of $n\%$ of company $x$, and a binary predicate *producedby*, where $producedby(a, x)$ tells that company $x$ produces good $a$.

Actions which can be performed are encoded within the *bought* predicate which indicates a given company has been bought (e.g. $bought(x)$ tells that $x$ has been bought), whereas effects show up on the *controlled* predicate symbol which encodes companies which are controlled (e.g. $controlled(x)$ tells that $x$ is controlled), and on the *produced* predicate which carries information about which goods are produced by the current set of controlled companies (e.g. $produced(a)$ means that in some way we control a company able to produce the good $a$).

---
[5]Note that the number of calls is polynomial in the input size, since $\log k$ is $O(|\mathcal{P}|)$ due to the succinct representation of numbers encoding penalties.

A suitable PAP $\mathcal{P} = \langle H, LP, O, \gamma, cost \rangle$ is built as follows. We set $H = \{bought(x_1), \ldots, bought(x_n)\}$ as the set of companies which can be hypothetically bought. For each atom $bought(x) \in H$, $\gamma(bought(x))$ is set to be the cost of acquiring company $x$, whereas $cost$ is the usual sum operator. We set $O = \{produced(y_1), \ldots, produced(y_n)\}$, where $y_1, \ldots, y_n$ is the set of goods to be produced. The program $LP$ is the following:

(1)  $produced(X) \leftarrow producedby(X, Y), controlled(Y).$
(2)  $controlled(X) \leftarrow bought(X).$
(3)  $controlled(X) \leftarrow share(X, Y, N), controlled(Y), N > 50.$
(4)  $controlled(X) \leftarrow share(X, Y, N), share(X, Y, M),$
     $\qquad\qquad\qquad\qquad controlled(Y), controlled(Z), M + N > 50.$

**Example 6.2 (Blocks world with penalization).** Planning is another scenario where abduction with penalization proves to be useful. We chosen a blocks world version where each block has a weight, and moving a block implies an expense proportional to block's weight. The goal is to discover cheapest plans among existing ones.

(1)  $location(B) \leftarrow block(B, \_).$
(2)  $location(T) \leftarrow table(T).$
(3)  $on(B, L, 0) \leftarrow init(B, L).$
(4)  $on(B, L, T1) \leftarrow block(B, \_), location(L), move(B, L, T), next(T, T1).$
(5)  $on(B, L, T1) \leftarrow on(B, L, T), next(T, T1), \neg moved(B, T).$
(6)  $moved(B, T) \leftarrow move(B, \_, T).$
(7)  $fail \leftarrow on(B, L, T), on(B, L1, T), L \neq L1.$
(8)  $fail \leftarrow block(B, \_), time(T), on(B, B, T).$
(9)  $fail \leftarrow move(B, L, T), move(B1, L1, T), B \neq B1.$
(10) $fail \leftarrow move(B, L, T), move(B1, L1, T), L \neq L1.$

Assume $B$ is the set of block available, $H$ is set to be $\bigcup_{b \in B, l \in L, t \in T} move(b, l, t)$ where $\gamma(move(b, l, t))$ is set to the weight of block $b$.

## 7  Conclusion

We have defined a formal model for abduction with penalization from logic programs. We have shown that the proposed formalism is highly expressive and it allows to encode relevant problems in an elegant and natural way. We have carefully analyzed the computational complexity of the main decisional problems arising in this framework. The complexity analysis shows some interesting properties of the formalism: (1) "negation comes for free", that is, the addition of (even unstratified) negation does not cause any further increase to the complexity of the abductive reasoning tasks (which is the same as for positive programs); (2) abduction with penalization over general logic programs has precisely the same complexity as abduction with penalization over definite Horn theories of classical logics. Consequently, the user can enjoy the knowledge representation power of nonmonotonic negation without paying any additional cost in terms of computational overhead. Moreover, the complexity analysis indicates that abduction with penalization has the same complexity as reasoning in (or-free) logic programming with weak constraints [2]. Future work indeed concerns the implementation of the proposed formalism on top of the DLV system [9] by exploiting weak constraints.

# References

[1] C. Baral and M. Gelfond. Logic Programming and Knowledge Representation *Journal of Logic Programming*, 1994.

[2] F. Buccafurri, N. Leone, and P. Rullo. Enhancing Disjunctive Datalog by Constraints. *IEEE Transactions on Knowledge and Data Engineering*, 12(5):845–860, 2000.

[3] E. Charniak and P. McDermott. *Introduction to Artificial Intelligence*. Addison Wesley, Menlo Park, Ca, 1985.

[4] L. Console, D. Theseider Dupré, and P. Torasso. On the Relationship Between Abduction and Deduction. *Journal of Logic and Computation*, 1(5):661–690, 1991.

[5] M. Denecker and D. De Schreye. Representing incomplete knowledge in abductive logic programming. In *Proc. of the International Symposium on Logic Programming*, pp. 147–163, 1993.

[6] P. Dung. Negation as Hypotheses: An Abductive Foundation for Logic Programming. In *Proceedings ICLP-91*. MIT Press, 1991.

[7] Eiter, T., Gottlob, G.. The Complexity of Logic-Based Abduction. *Journal of the ACM*, 42(1):3–42, 1995.

[8] Eiter, T., Gottlob, G., and Mannila, H.. Disjunctive Datalog. *ACM Transactions on Database Systems*, 22(3):364–418, 1997.

[9] T. Eiter, W. Faber, N. Leone, and G. Pfeifer. Declarative Problem-Solving Using the DLV System. *Logic-Based Artificial Intelligence*. Kluwer Academic Publishers, 2000.

[10] M. Gelfond and V. Lifschitz. The Stable Model Semantics for Logic Programming. In *Proc. Fifth Logic Programming Symposium*, pp. 1070–1080. MIT Press, 1988.

[11] J. R. Hobbs and M. E. Stickel. Interpretation as Abduction. In *Proc. 26th Annual Meeting of the Assoc. for Computational Linguistics*, 1988.

[12] A. Kakas and R. Kowalski and F. Toni. Abductive Logic Programming. *Journal of Logic and Computation*, 2(6):719–771, 1992.

[13] A. Kakas and P. Mancarella. Generalized Stable Models: a Semantics for Abduction. In *Proc. of ECAI-90*, pp. 385–391, 1990.

[14] A. Kakas and P. Mancarella. Database Updates Through Abduction. In *Proceedings VLDB-90*, pp. 650–661, 1990.

[15] K. Konolige. Abduction versus closure in causal theories. *Artificial Intelligence*, 53:255–272, 1992.

[16] Papadimitriou, C.. The Complexity of Unique Solutions. *Journal of the ACM* 31, 492–500, 1984.

[17] C. H. Papadimitriou. *Computational Complexity*. Addison-Wesley, 1994.

[18] C. S. Peirce. Abduction and induction. In J. Buchler, editor, *Philosophical Writings of Peirce*, chapter 11. Dover, New York, 1955.

[19] D. Poole. Normality and Faults in Logic Based Diagnosis. In *Proceedings IJCAI-89*, pp. 1304–1310, 1989.

[20] C. Sakama and K. Inoue. On the Equivalence between Disjunctive and Abductive Logic Programs. In *Proc. of ICLP-94*, pp. 88–100, 1994.